# Ultra-low Thermal Conductivity of Nanogranular Indium Tin Oxide Films Deposited by Spray Pyrolysis

Vladimir I. Brinzari[1], Alexandr I. Cocemasov[1], Denis L. Nika[1,*], Ghenadii S. Korotcenkov[2]

[1] E. Pokatilov Laboratory of Physics and Engineering of Nanomaterials, Department of Theoretical Physics, Moldova State University, Chisinau, MD-2009, Republic of Moldova

[2] Gwangju Institute of Science and Technology, Gwangju 500-712, Republic of Korea

**Abstract**

The authors have shown that nanogranular indium tin oxide (ITO) films, deposited by spray pyrolysis on silicon substrate, demonstrate ultralow thermal conductivity $\kappa \sim 0.84 \pm 0.12$ Wm$^{-1}$K$^{-1}$ at room temperature. This value is approximately by one order of magnitude lower than that in bulk ITO. The strong drop of thermal conductivity is explained by nanogranular structure and porosity of ITO films, resulting in enhanced phonon scattering on grain boundaries. The experimental results were interpreted theoretically, employing Boltzmann transport equation approach for phonon transport and filtering model for electronic transport. The calculated values of thermal conductivity are in a reasonable agreement with the experimental findings. Our results show that ITO films with optimal nanogranular structure may be prospective for thermoelectric applications.

* Corresponding author: dlnika@yahoo.com (D.L. Nika).

## 1. Introduction

Thermoelectric materials are described by a figure of merit *ZT*, which reveals how well a material converts heat into electrical energy:

$$ZT = \frac{\sigma S^2 T}{\kappa} \qquad (1)$$

where $\sigma$, $S$, $\kappa$, and $T$ are the electrical conductivity, Seebeck coefficient or thermopower, thermal conductivity and absolute temperature, respectively. Electrical part of the heat conversion is usually characterized by a power factor $PF = \sigma S^2$.

*ZT* is directly related to the efficiency of thermoelectric energy conversion - the higher *ZT* means the better efficiency. In order to enhance *ZT* of a material one needs to increase its power factor and simultaneously decrease its thermal conductivity. Since these transport characteristics depend on interrelated material properties, a set of parameters such as carrier concentration, effective mass, electronic and lattice thermal conductivity ~~need to be mutually~~ should be mutually optimized [1]. The power tool for the optimization of thermal transport at nanoscale is phonon engineering [2-4].

Indium tin oxide is a metal oxide, which consists of a mixture of indium (III) oxide $In_2O_3$ and tin (IV) oxide $SnO_2$. ITO is a highly degenerate *n*-type semiconductor with a wide optical band gap of 3.5 – 4.3 eV [5]. Among the most studied conductive oxides ITO possesses a large variety of unique properties like high electrical conductivity, high optical transparency in the visible region, high reflectance in the infrared spectral regions, high thermal stability, good adhesion to the substrate and ease of patterning to form transparent electrodes [6]. The maximal values of electrical conductivity and optical transparency were reported for ITO with 10% of Sn [7]. Such kind of ITO is the industry standard in transparent conducting oxides (TCO) and is heavily exploited in different practical applications [5, 8].

Owing to ionic nature, conductive metal oxides like ITO tend to have carrier mobility by an order of magnitude lower than Si and other covalent compounds [9]. Moreover, large bonding energies of the ionic bonds and the small atomic mass of oxygen results in a high velocity of phonon waves propagating through the crystal lattice of the oxide compound. Therefore bulk-like ITO with tightly-packed large grains (~5 μm) possesses relatively high total thermal conductivity $\kappa \sim 10$ Wm$^{-1}$K$^{-1}$ [10, 11] at room temperature (RT).

Due to their low carrier mobility and high thermal conductivity metal oxides have been ignored as potential thermoelectric materials until the beginning of 1990s, however the theoretical predictions, suggesting that the thermoelectric efficiency could be greatly enhanced with the use of nanoscale engineering began to appear in scientific reports [12].

Measurements of $In_2O_3$-based compounds confirm the rapid decrease of thermal conductivity with the reduction of grain size in comparison with bulk values [13, 14]. This effect is due to reinforcement of phonon scattering on grains and nano-inclusions, leading to a glass-like thermal conductivity. Lan et al. reported on achievement of thermal conductivity below the amorphous limit $\kappa \sim 1.2$ $Wm^{-1}K^{-1}$ at 973 K in sintered $In_2O_3$ doped by Zn and Ce [11]. Creation of the porous structure is another key possibility to strongly reduce the thermal conductivity. M. Ohtaki et al. [15, 16] demonstrated that in $In_2O_3 \cdot MO_x$ (M=Cr, Mn, Ni, Zn, Sn) ceramics the thermal conductivity decreases to $\kappa \sim 1.58 - 1.75$ $Wm^{-1}K^{-1}$ at RT irrespective of chemical nature of the second phase. The relative density of prepared compounds was around 65-70 %. Although it was not estimated in Ref. [15], their samples can be attributed to micro- mesoporous ceramics (in terms of International Union of Pure and Applied Chemistry (IUPAC) typical pore scale is between 2-50 nm).

Several independent recent studies [14, 17-24] confirm the prospective of ITO and $In_2O_3$-based compounds for the thermoelectric energy conversion due to the enhancement of power factor. It was demonstrated that *PF* is strongly affected by Sn content and nanogranular structure. However the thermal conductivity properties of nanogranular ITO remain largely unexplored. In this letter we investigate the thermal conductivity of nanogranular ITO films, deposited by spray pyrolysis technique. We show that thermal conductivity is strongly suppressed due to enhancement of phonon scattering on grain boundaries and reaches ultra-low values $\sim 0.84 \pm 0.12$ $Wm^{-1}K^{-1}$ at room temperature.

ITO film deposition on polished alumina and silicon substrates was performed by spray pyrolysis method [17]. For this purpose a mixture of 0.2 M water solution of $InCl_3$ and $SnCl_4 \cdot 5H_2O$ was used for deposition of thin films with different thickness (in the range of 100-400 nm) and Sn/In ratios at pyrolysis temperature of 350 $^o$C. Silver paste electrodes were applied to samples with films deposited on the alumina substrates. All experimental samples were annealed for stabilization of film's parameters at $T_{an}$=550 $^o$C for 0.5h.

The samples were characterized using the scanning electron microscopy (SEM), Micro-

Raman spectroscopy and energy-dispersive x-ray spectroscopy (EDS). SEM images were taken on Leo SUPRA 55 SEM/E-beam lithography system. Micro-Raman spectroscopy was performed on a Renishaw InVia instrument in the backscattering configuration under λ = 488 nm and 633 nm laser excitation, respectively. The excitation laser power varied from 1 to 10 mW. Elemental analysis of samples was done by energy dispersive X-ray spectroscopy in the XL30 SEM fitted with EDAX® Genesis system. LayerProbe software package from Oxford Instruments was used to calculate the thickness of our films. The approach uses the generated X-ray intensities and calculates the absorption to determine the layer thicknesses. Based on the spectra taken at 5 different spots of each sample, the thicknesses of our samples were evaluated.

The measurements of the cross-plane thermal conductivity of samples deposited on silicon substrate were performed using the 'laser flash' method (Netzsch LFA) in conventional configuration. The Laser Flash technique (LFT) is a transient method that directly measures the thermal diffusivity, α. To perform LFT measurement, each sample was placed into a sample holder that fitted its size. The bottom of the sample was illuminated by a flash of a xenon lamp and the temperature of the opposite surface of the sample was monitored with a cryogenically cooled InSb infra-red (IR) detector. The temperature rise as a function of time, △T(t) was used to extract α. In order to extract the specific heat capacity, $C_p$, comparison of △T(t) of the sample to that of a reference sample (bulk Si) was performed under the same experimental conditions. Thermal conductivity of sandwich structure was determined from the equation $\kappa = \rho \alpha C_p$, where ρ is the mass density of the material.

Raman spectra have shown that ITO vibrational modes are close to same ones of $In_2O_3$ [25] at Sn doping less than 10 at. %. They reveal up to 10 peaks originated from light scattering on crystalline lattice that belongs to $Ia\overline{3}/T_h^7$ space group (see Fig. 1). At the same time, faint peaks at 162, 240, 433, 474, 585 and 633 $cm^{-1}$, which correspond to Sn – O vibrational modes [26, 27], can be also resolved. EDS data confirm the same Sn/In content in sprayed solutions and deposited films. Granular nanostructure of the films is sensitive enough to Sn content. We observed a sharp decrease of grain size and changes in grain shape in the range of 0-10% of Sn (see Fig. 2 (a,b)). Specifically, inside this range of Sn content, i.e. near 5% there was found maximal thermoelectric PF for our films [17, 28]: PF ~ 4 mW/(m $K^2$) at T~$300^0$C, which is by a factor of 4-5 higher than that in conventionally prepared ITO (10 % Sn and micrometer grain sizes). At this composition ITO films demonstrate grain size growth on film's thickness (see Fig. 2 (a, c, d)). Previous studies [17] have shown that films with given optimal composition are textured with preferential surface orientation in direction (100). SEM images, corresponding to ITO with ~5% of Sn, demonstrate in-plane orthogonal or square shapes of grains in agreement with this finding.

The measurement of cross-plane thermal conductivity was performed for bare Si substrate with thickness $d_{Si}$ = 480 μm and three different ITO/Si sandwich samples with total thickness $d_j = d_{Si} + d_{ITO,j}$, where $d_{ITO,j}$ is the thickness of ITO film: $d_{ITO,1}$ =90 nm (sample 1), $d_{ITO,2}$ =170 nm (sample 2) and $d_{ITO,3}$ =340 nm (sample 3). The dependence of cross-plane thermal conductivity on temperature in Si substrates and ITO/Si samples (S-1), (S-2) and (S-3) is presented in Fig. 3(a). Thermal conductivity of ITO is lower than that in bulk Si, therefore rise of $d_{ITO}$ decreases the thermal conductivity of ITO/Si samples.

The thermal conductivity in ITO/Si samples can be described as [29]

$$\frac{d_j}{\kappa_j} = \frac{d_{Si}}{\kappa_{Si}} + \frac{d_{ITO,j}}{\kappa_{ITO}} + R_b, \quad j = 1,2,3 \qquad (2)$$

where $\kappa_j$ is the thermal conductivity of ITO/Si sample (S-j) with the thickness $d_j$ and $R_b$ is the thermal boundary resistance (TBR) between ITO film and Si substrate. Since $d_{ITO} \ll d_{Si}$ for all samples one can extract $\kappa_{ITO}$, taking into consideration three different pairs of Eqs. (2): (j=1, j=2); (j=1, j=3) and (j=2, j=3):

$$\kappa_{ITO} = \kappa_{mn} = \frac{\kappa_m \kappa_n (d_n - d_m)}{d_{Si}(\kappa_m - \kappa_n)}, \quad m = 1,2,3; \ m < n \leq 3. \qquad (3)$$

The dependencies of thermal conductivities of ITO $\kappa_{12}$, $\kappa_{13}$ and $\kappa_{23}$ on the temperature are shown in Fig. 3(b). The thermal conductivity weakly decreases with temperature rise up to 140°C. Averaging $\kappa_{12}$, $\kappa_{13}$ and $\kappa_{23}$ we estimated the value of RT thermal conductivity in ITO as $\kappa_{ITO}$=0.82±0.12 Wm$^{-1}$K$^{-1}$. The average RT thermal boundary resistance is ~2·10$^{-7}$ m$^2$KW$^{-1}$. The variation in thermal conductivity values in Fig. 3(b) is explained by different dispersion of grain sizes in S-1, S-2 and S-3 samples, resulting in different strength of phonon scattering on the grains. However it is difficult to predict the dependence of $\kappa_{ITO}$ on $d_{ITO}$ from our measurements because size, stacking order and mutual orientations of grains simultaneously change with film thickness. These geometrical factors also affect the effective cross section and porosity of the film, and as a result determine the thermal conductivity. Data scattering for $R_b$ is smaller because it is determined mostly by the quality of ITO/Si interfaces.

Obtained values of RT $\kappa_{ITO}$ is approximately by one order of magnitude lower than that of bulk ITO [10, 11] and by a factor of 2-5 lower than that of ITO thin films prepared by other techniques [13, 15-16]. Li et al. [30] has reported that large dispersion of grain size in nanostructured films and inevitable appearance of porosity should lead to enhancement of phonon scattering and reduction of thermal conductivity. Such special structural properties are inherent for films deposited by spray pyrolysis. Our results for ITO films confirm this reasoning. Possible anisotropy of in-plane and cross-plane thermal conductivities in ITO films may be expected due to anisotropy of crystallites growth, their specific shapes and contacts between them. The investigations of possible thermal conductivity anisotropy require further separate study.

For the interpretation and validation of our experimental results we performed theoretical calculations of thermal conductivity in nanogranular ITO films. In the framework of Boltzmann transport equation approach, the phonon thermal conductivity $\kappa_{ph}$ in ITO film is given by

$$\kappa_{ph}^{3D} = \frac{\hbar^2 \langle \upsilon \rangle^2}{2\pi^2 k_B T^2} \int_0^{2\pi/a} \omega^2(q)\tau(q) \frac{\exp(\hbar\omega(q)/k_B T)}{[\exp(\hbar\omega(q)/k_B T)-1]^2} q^2 dq, \qquad (4)$$

where $\langle \upsilon \rangle$=4.35 km/s is the average sound velocity in bulk ITO, $a$ = 1.012 nm is the lattice constant, $q$ is the magnitude of the phonon wave vector, $\omega = \langle \upsilon \rangle q$ is the phonon frequency, $\tau(q)$ is the total phonon relaxation time, $T$ is the absolute temperature, $k_B$ and $\hbar$ is the Boltzmann's and Plank's constant, respectively. The total phonon relaxation time was calculated using Matthiessen's rule as $1/\tau = 1/\tau_U + 1/\tau_G$, where $1/\tau_U(q) = \alpha T \omega^2(q) \exp[-T^{-1}]$ is the three-phonon Umklapp scattering rate and $1/\tau_G = \langle \upsilon \rangle / l_G$ is the phonon scattering rate on grain boundaries with the grain size $l_G$. Parameter $\alpha$ was obtained from the comparison between calculated and measured [10] thermal conductivities in bulk ITO (without phonon scattering on grains): $\alpha = 2.45 \times 10^{-18}$ s/K. We assumed here that for our films with $\sigma \sim 10^5$ S/m phonon thermal conductivity constitutes 2/3 [13] of the total thermal conductivity measured in Ref. [10]. The Eq. (4) describes the heat flux carrying by the phonons with three-dimensional density of states. However in thin films the phonon density of states (PDOS) changes from three- to two-dimensional (2D). To take into account possible change of PDOS in ITO films we also estimated 2D phonon thermal conductivity from the following equation:

$$\kappa_{ph}^{2D} = \frac{\hbar^2 \langle \upsilon \rangle^2}{4\pi k_B T^2} \frac{N_b}{d} \int_0^{2\pi/a} \omega^2(q)\tau(q) \frac{\exp(\hbar\omega(q)/k_B T)}{[\exp(\hbar\omega(q)/k_B T)-1]^2} q dq, \qquad (5)$$

where $d$ is the film thickness. Number of acoustic phonon branches $N_b$ depends on the film thickness and can be roughly estimated as $N_b = 3 \times d/(2 \times a)$ [31].

Electronic thermal conductivity $\kappa_{el}$ was calculated in the framework of a so-called filtering model (FM) [32, 33] developed recently by some of us for granular ITO films [34]:

$$\kappa_{el} = \frac{4N_c k_B^2 T}{3m^* \sqrt{\pi}} \int_{\varepsilon_b}^{\infty} \tau(E) \varepsilon^{3/2} (\varepsilon - \varepsilon_F)^2 \frac{e^{\varepsilon - \varepsilon_F}}{(1+e^{\varepsilon-\varepsilon_F})^2} d\varepsilon - \sigma S^2 T \qquad (6)$$

where: $\varepsilon = E/k_B T$, $\varepsilon_F = E_F/k_B T$, $\varepsilon_b = E_b/k_B T$ and $N_c = 2\left(\frac{m^* k_B T}{2\pi\hbar^2}\right)^{3/2}$. Here $E$ is the electron energy, $E_b$ is the height of potential barrier, formed in the vicinity of the grain boundaries, $E_F$ is the Fermi energy, $\tau(E)$ is the electron momentum relaxation time, $m^*$ is the electron effective mass and $N_c$ is the effective density of states in conduction band. Fundamental parameters of ITO used in the calculations were taken from Ref. [9]. Two dominant mechanisms of electron scattering were considered: scattering by polar optical phonons and ionized impurities. The calculation of values $\sigma$ and $S$ was also carried out within the FM [34].

The dependence of phonon thermal conductivity $\kappa_{ph}^{3D}$ and $\kappa_{ph}^{2D}$ on grain size is presented in Fig. 4(a) for different temperatures. The thermal conductivity drops by a factor of ~ 2 with decrease of the grain size from 100 nm to 20 nm. Temperature dependence of phonon, electron and total thermal conductivity is presented in Fig. 4(b) by red, magenta and black curves, correspondingly. The phonon thermal conductivities are plotted for grain size $l_G$=20 nm which is typical for our ITO films. The phonon thermal conductivity decreases with temperature due to an enhancement of phonon-phonon scattering, while electronic thermal conductivity demonstrates opposite trend. The non-linear growth of $\kappa_{el}$ is explained by increase of total number of electrons participating in electronic transport over the barrier with temperature rise.

The calculated RT total thermal conductivity $\kappa$=2.5 Wm$^{-1}$K$^{-1}$ is larger than average experimental value. This discrepancy may be attributed to two reasons: (i) the porosity of ITO films and (ii) anisotropy in grains size, shape and alignment. Using effective medium approximation [35, 36], we have estimated the effect of the porosity: 30% porosity decreases the thermal conductivity by 39 % for 3D phonons and by 57 % for 2D. The obtained RT 2D thermal conductivity $\kappa_{ph}^{2D} \sim 1.1$ W/mK is in a reasonable agreement with the experimental value $\kappa_{ITO}$=0.82±0.12 Wm$^{-1}$K$^{-1}$. Slightly lower values of experimental $\kappa_{ITO}$ can be attributed to irregularity and scattering in size, shape and orientation of grains, resulting in possible trapping of phonon modes in grain segments. This effect is similar to phonon modes trapping in segmented and cross-section modulated nanowires, leading to the drastic reduction of phonon thermal conductivity [37-39].

The measurements of electrical conductivity, Seebeck coefficient and power factor in ITO films deposited by spray pyrolysis was recently reported by some of us in Ref. [17, 28]. Using $PF\sim3$ mW/(m K$^2$) at $T\sim300$ °C, obtained in thermal stability experiments [28] and upper value of measured $\kappa_{ITO}$= 0.9 Wm$^{-1}$K$^{-1}$ from this work we can roughly estimate $ZT \sim 1.9$ for optimally doped ITO films. These unusually high values of ZT are related to simultaneous improvement of two main thermoelectric parameters: Seebeck coefficient and thermal conductivity. Proper selection of technological conditions of the film growth allows one to achieve nanogranular structure providing optimal structural (grain size, grain stacking) and electro-physical parameters (Sn distribution inside the grain, and a potential barrier in the vicinity of grain boundary).

In conclusion, we have shown that ITO films deposited by spray pyrolysis possess ultralow room temperature thermal conductivity $\kappa \sim 0.84\pm0.12$ Wm$^{-1}$K$^{-1}$, which is by an order of magnitude lower than that in bulk ITO. The drop of thermal conductivity is explained by nanogranular structure of ITO films with average grains size ~ 20 nm, resulting in strong phonon scatterings on grains. The ultralow thermal conductivity combined with high enough power factor may lead to thermoelectric application of ITO films prepared by spray pyrolysis.

**Acknowledgements**

AIC and DLN thank Prof. Alexander A. Balandin for his kind hospitality during their work in the Nano-Device Laboratory at the University of California – Riverside, and for providing the results of the thermal conductivity measurements conducted in his laboratory. VIB, AIC and DLN acknowledge the financial support from the Republic of Moldova through the project 15.817.02.29F. The work in Republic of Korea was supported in part by the Ministry of Science, ICT and Future Planning (MSIP) and by the National Research Foundation (NRF)

grants No. 2011-0028736 and No. 2013-K000315.

**FIGURE CAPTIONS**

**Figure 1.** Raman spectrum of ITO film with 5% of Sn deposited by spray pyrolysis on Si substrate.

**Figure 2.** SEM images of nanogranular ITO films, deposited by spray pyrolysis on silicon substrate, shown for: (a) 90 nm - thick ITO film with 5% of Sn; (b) 90 nm - thick ITO with 10% of Sn; (c) 170 nm - thick ITO film with 5% of Sn and (d) 340 nm – thick ITO film with 5% of Sn.

**Figure 3.** (a) Measured cross-plane thermal conductivity as a function of temperature for ITO films (5% of Sn) on Si substrate with different film thicknesses: 90 nm (*S-1*), 170 nm (*S-2*) and 340 nm (*S-3*). The results for Si substrate without ITO film are also shown by black squares. (b) The dependence of cross-plane thermal conductivity of ITO films on the temperature. The thermal conductivity data were extracted using Eq. (3).

**Figure 4.** (a) Thermal conductivity as a function of average grain size plotted for different temperatures. Dashed curves show the thermal conductivity calculated with taking into account 2D PDOS (see Eq. 5). (b) Total thermal conductivities as a function of the temperature plotted for different porosity of ITO films. The phonon (red curve) and electronic (magenta curve) thermal conductivities are also presented. The results are shown for average grain size $l_G$ = 20 nm and Fermi energy $E_F$ = 0.35 eV.

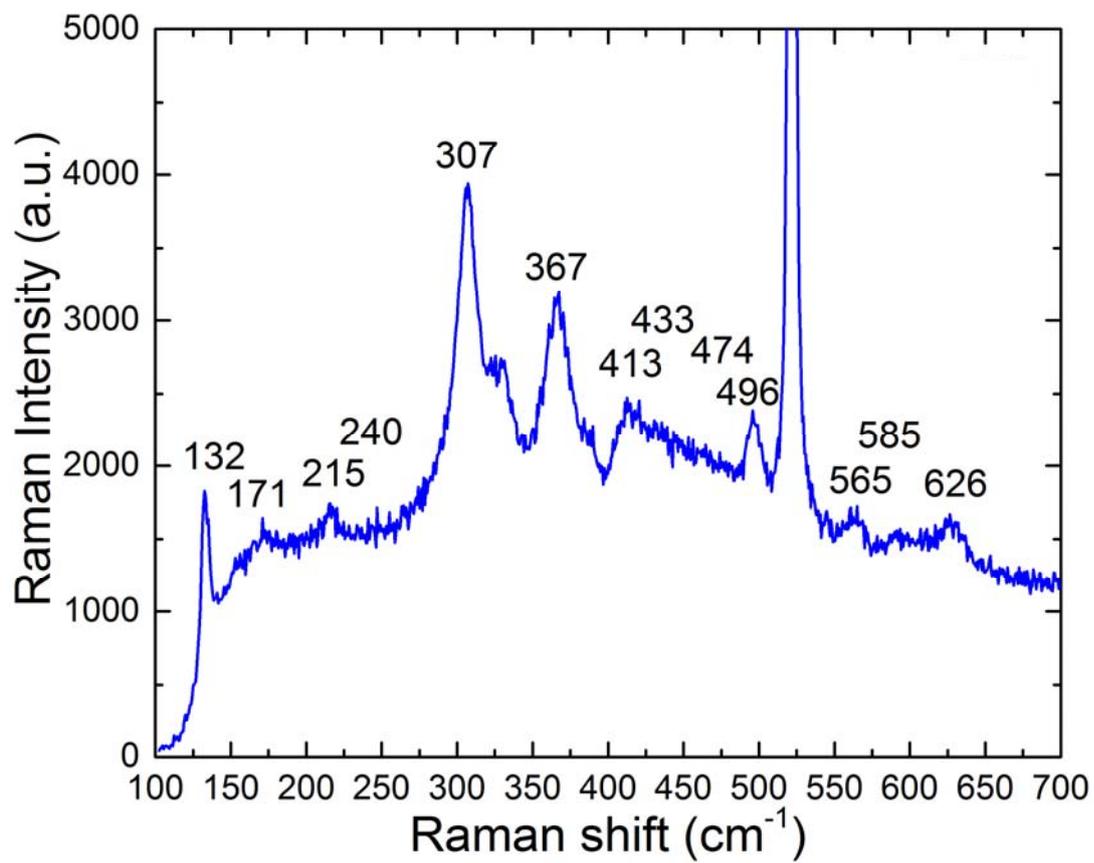

**Figure 1 of 4**. V. Brinzari et al.

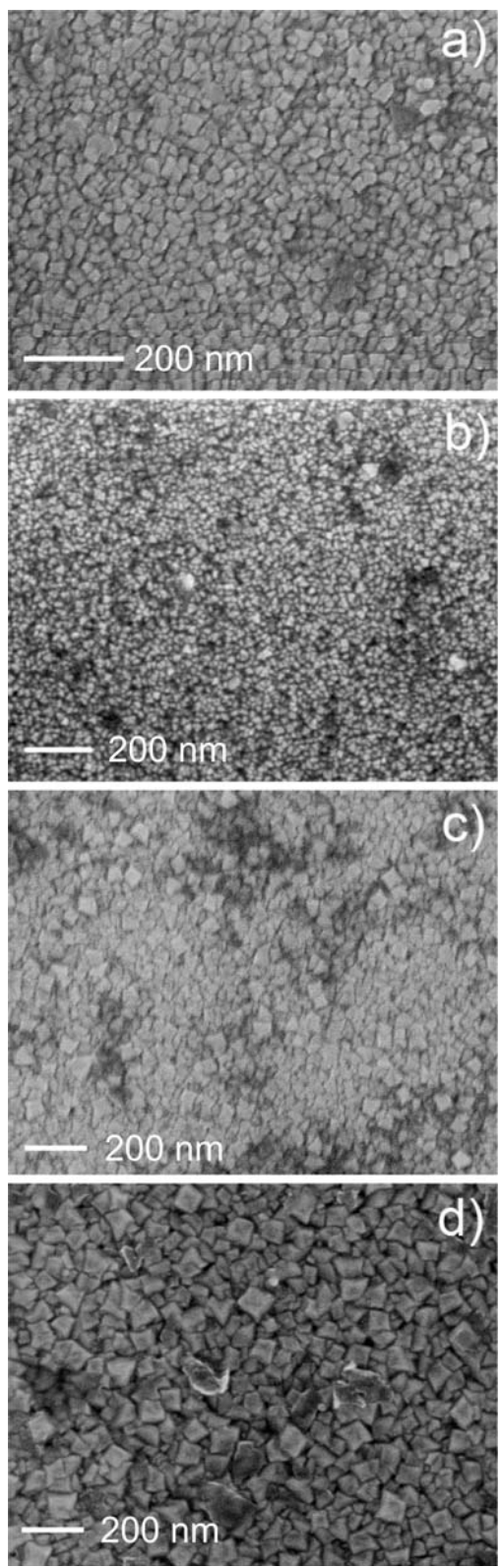

**Figure 2 of 4**. V. Brinzari et al.

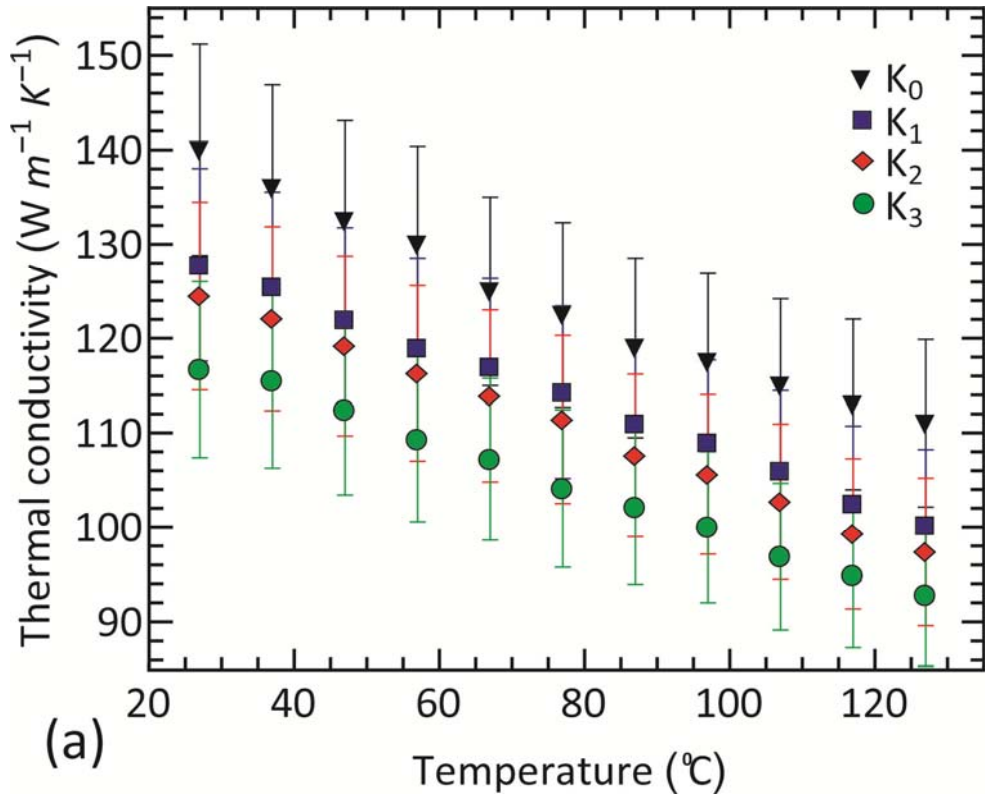

(a)

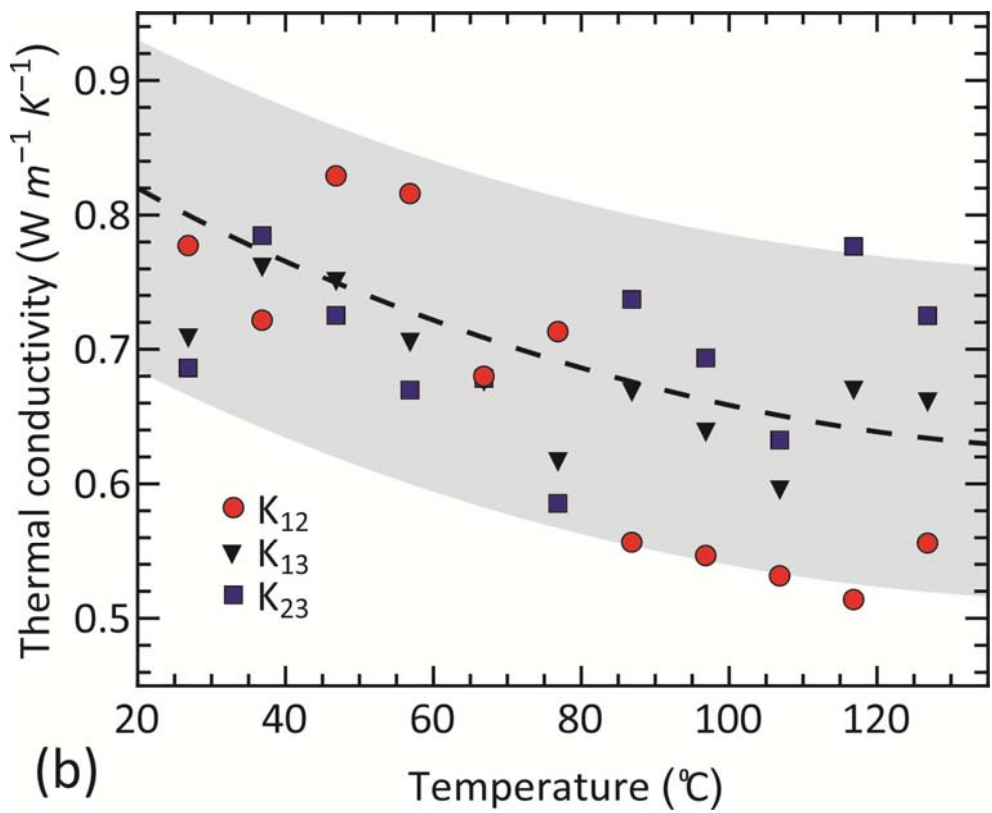

(b)

**Figure 3 of 4**. V. Brinzari et al.

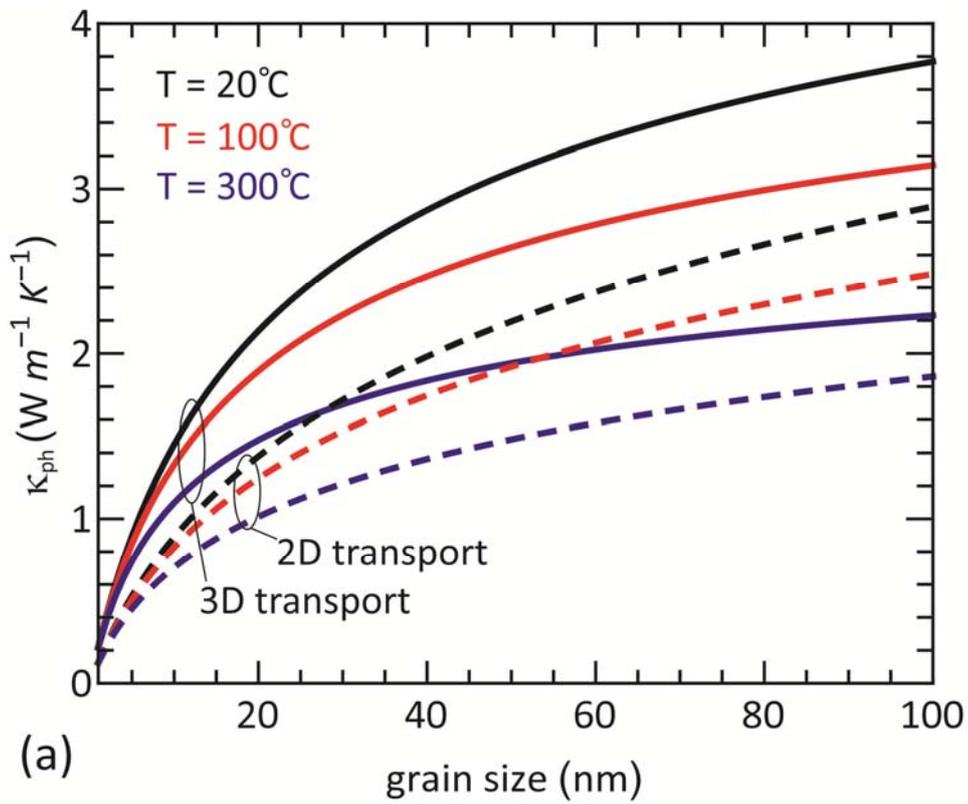
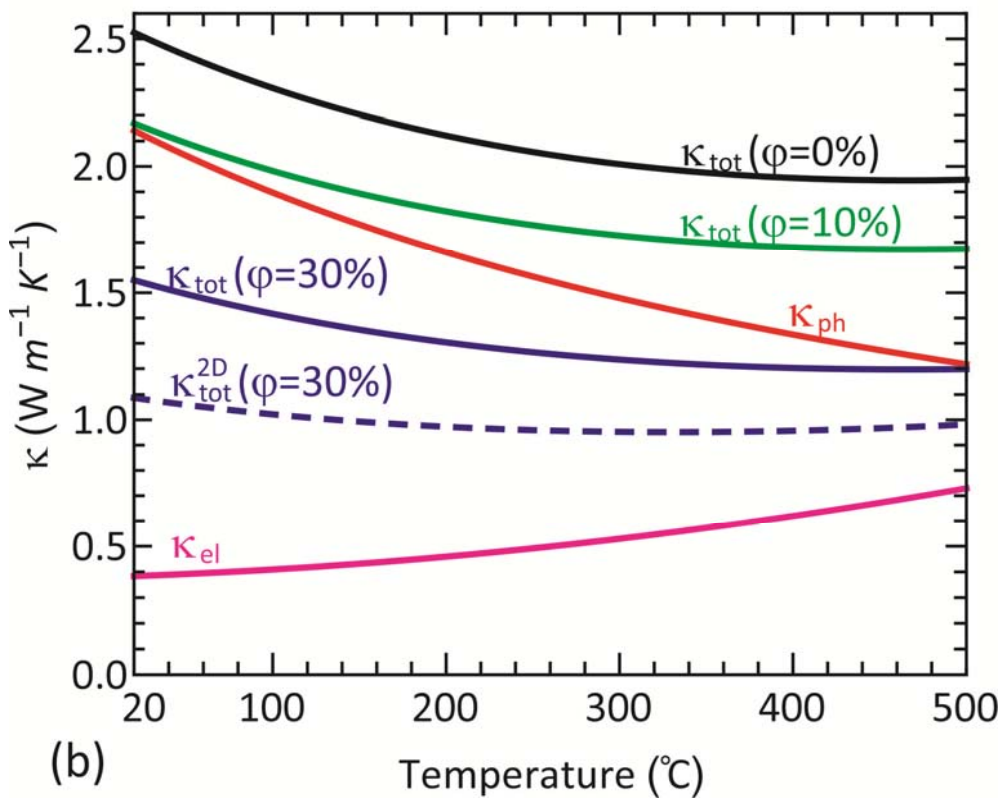

**Figure 4 of 4**. V. Brinzari et al.